\documentclass[traditabstract]{aa} 
\usepackage{graphicx}
\usepackage{txfonts}
\begin{document}
\title{What is needed to accept the new explanation of DAMA results}

\author{J.Va'vra
\inst{}
}

\institute{SLAC, Stanford University, CA94309, U.S.A.\\
\email{jjv@slac.stanford.edu}
}

\date{Received January 25, 2014}


\abstract
{The DAMA experiment clearly observes a small oscillatory signal. The observed yearly modulation is in phase with the Earth's motion around the Sun.   
Recent reference [Vavra, 2014] suggested that the DAMA experiment observes a WIMP of much smaller mass than what Xenon 10, Xenon 100, LUX and CDMS experiments can possibly reach. 
Scattering would occur on proton or oxygen target present in the NaI(Tl) crystal as OH-contamination at a few ppm level. This paper elaborates further on the idea 
that the OH-molecule could act as a very sensitive detection mechanism for neutrons or WIMPs, and suggests a calibration procedure to prove this idea. 
We also propose a new detector concept to detect a low mass WIMP.} 


\maketitle
%

\section{Introduction}

The fact that the DAMA's result [Bernabei, 2013] is not confirmed by other Dark Matter searching experiments, such as CDMS, Xenon-10, Xenon-100 or LUX, gave the author 
an idea that the DAMA signal is due to scattering of a low mass WIMP with proton of hydrogen atom. The hydrogen is 
present in a form of small OH-contamination in NaI(Tl) crystals. OH-molecules may come from the primary NaI salt and could 
be present at a level of $\sim$few~ppm. The important point to realize is that the OH molecule would be sensitive to very low energy neutrons and low mass 
WIMPs, i.e.,  collision of the WIMP and proton will cause vibrations of this molecule, which could be detected.

The question is what is needed to prove that this hypothesis is correct ?

\begin{figure}[tbp]
\includegraphics[width=0.5\textwidth]{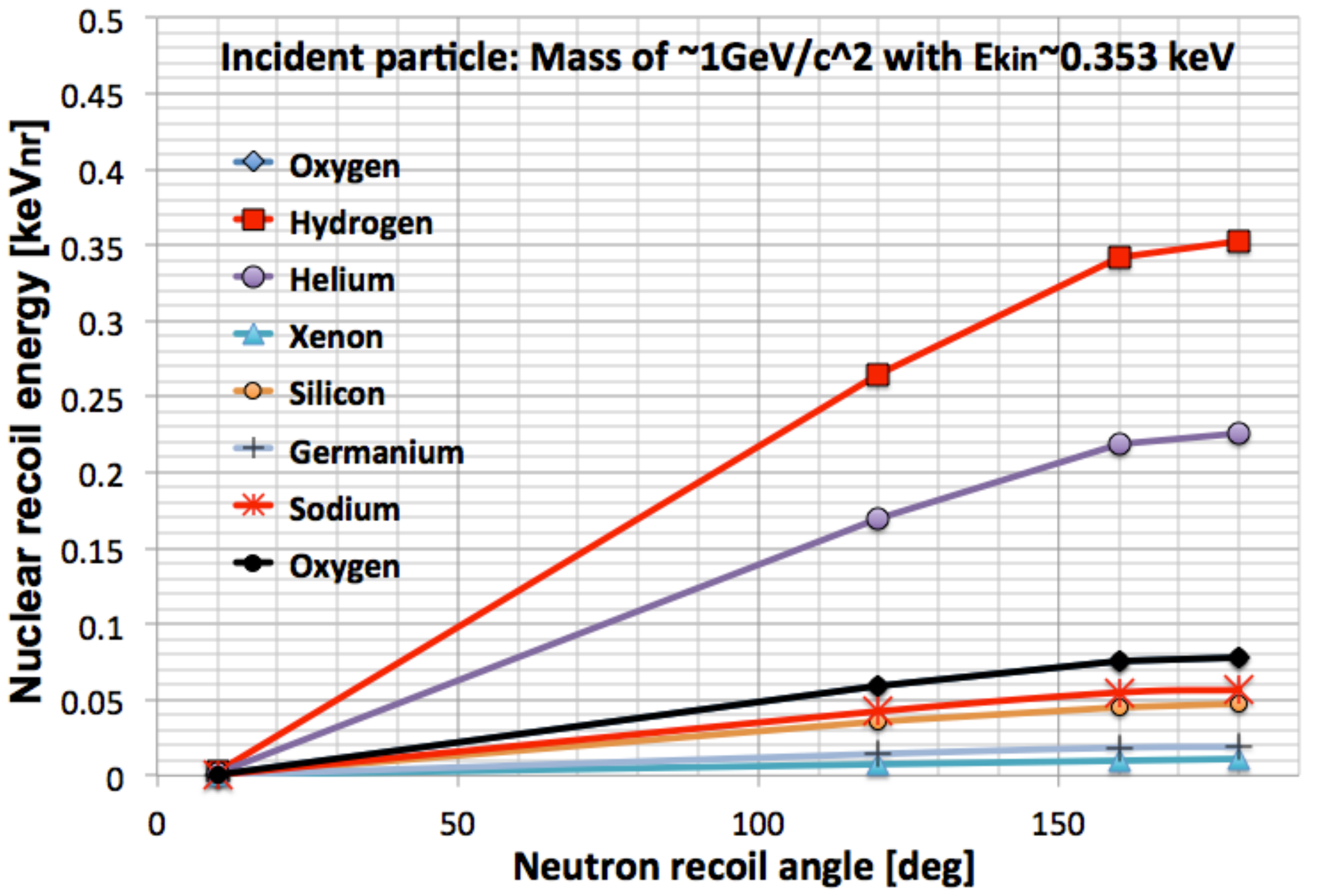}
\caption{Nuclear recoil energies for various nuclei assuming a $\sim$1~GeV/$c^2$ WIMP with kinetic energy of $\sim$0.353keV.}
\label{fig:Nuclear_recoil_from_1GeV}
\end{figure}

A low mass WIMP, for example $\sim$1~GeV/$c^2$, represents a real experimental challenge, as it requires a low mass target and extremely sensitive detector. 
The Dark Matter cloud is believed to be stationary relatively to the Galagxy. The Earth moves with a velocity of 230~+-~30~km/sec in the Galactic plane, 
and as a result, a $\sim$1~GeV/$c^2$ mass WIMP's kinetic energy oscillates between $\sim$0.353~keV and $\sim$0.208~keV relative to the DAMA 
experiment, on its yearly journey around the Sun. Figure~\ref{fig:Nuclear_recoil_from_1GeV} shows nuclear recoil energies as a function of recoil angle for various nuclei, 
assuming the WIMP mass of 1~GeV/$c^2$ WIMP. It is clear, that one prefers to use the proton target if the light WIMP mass is this small. 

\section{OH-impurity as a detector of WIMP}

\begin{figure}[tbp]
\includegraphics[width=0.5\textwidth]{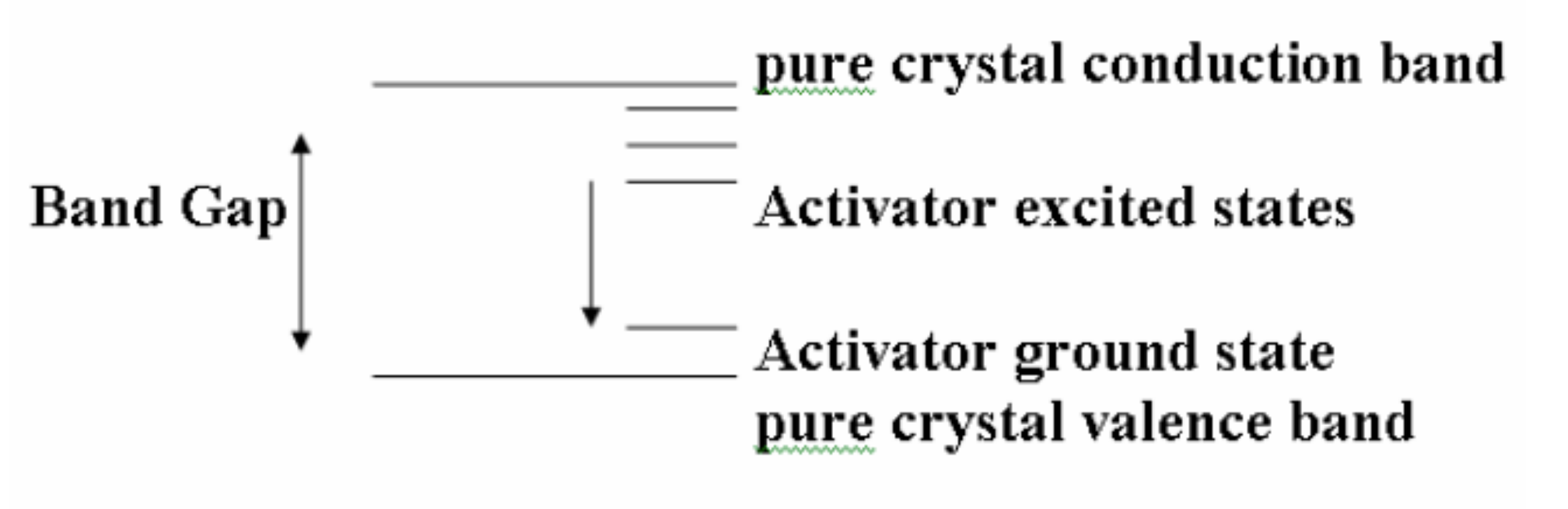}
\caption{Energy levels of NaI crystal with Tl-activator [Knoll, 2010].}
\label{fig:NaI_energy_levels}
\end{figure}

Fig.~\ref{fig:NaI_energy_levels} shows a classical model of the signal formation in NaI(Tl) crystals. A deposited energy creates electron-hole pair, excites an 
electron into the conduction band, where it moves until it finds an activator (Tl), with a very low ionization potential of 6.108~eV, where the 
de-excitation occurs via small photonic emissions, mostly in visible spectrum [Knoll, 2010]. 

\begin{figure}[tbp]
\includegraphics[width=0.5\textwidth]{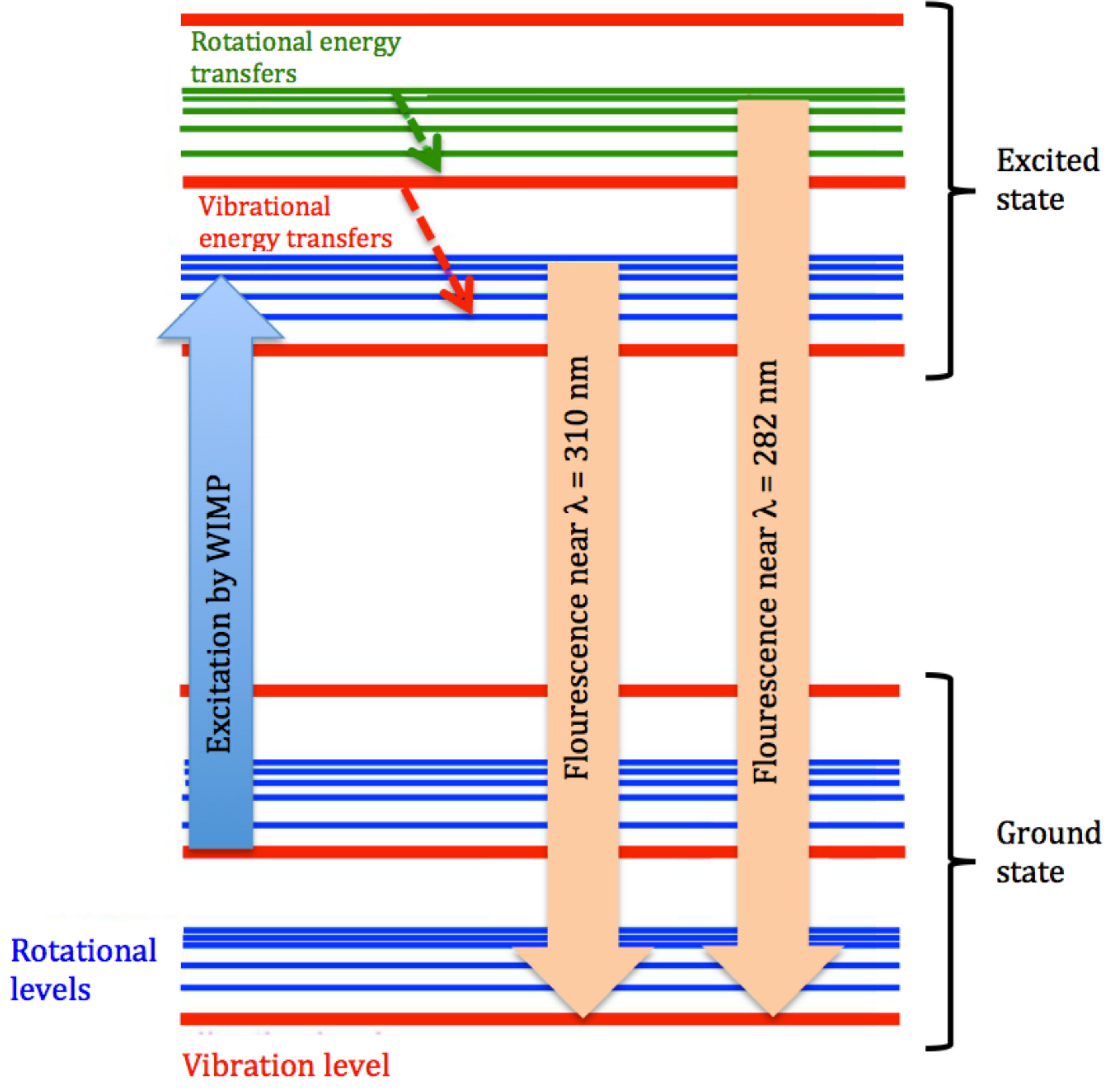}
\caption{Energy levels of OH-molecule are very complex. A low mass WIMP hits proton or even oxygen nucleus in the OH-molecule, causing excitation and subsequent 
photon fluoroscence emissions near 310 or 282~nm, which are detectable by the Bialkali photocathode, if one uses the right optical coupling allowing tranmsision of these wavelengths. 
In principle a very low mass WIMP could be detected by this technique, as 282~nm corresponds to only 4.67~eV.}
\label{fig:OH_levels}
\end{figure}

In this section we describe in more deatil what are possible excitation of the OH-molecule [Vavra, 2014] and answer a question if resulting photons are detectable. 
Fig.~\ref{fig:OH_levels} shows schematically energy levels in OH-molecule. If either a neutron or low mass WIMP hits a proton, or even an oxygen nucleus,
in the OH molecule, it will cause vibrations and molecular excitations to higher energy levels. The molecule can then de-excite by fluorescent 
photons at either $\sim$282~nm or $\sim$310~nm. Such photons can be detected by the Bialkali photocathode in principle, if one choses the optical coupling correctly. 
Through this mechanism one could then increase the sensitive to extremely low sub-keV energy deposits. 

One should mention that the OH-molecule was studied extensively by laser-induced fluorescence by many chemists. An example of such fluorescence measurement is 
the OH-molecule excitation by a 282~nm dye laser and in turn observing the 310~nm wavelength with a PMT with a notch filter [Smith, 1990]. 
They used this method to determine traces of OH-radicals in atmosphere [Matsumi, 2002].

One should also point out that 282~nm corresponds to 4.67~eV energy, which is a very small excitation compared to usual energy needed to excite NaI(Tl) crystal 
($\sim$25eV per electron-hole pair on average). This point plus a sensitivity to slow neutrons makes the OH-molecule excitation a very attractive idea 
to Dark Matter detectors. Just as one can pump the OH-molecule to excitation with a laser, one can excite it with WIMP collisions.

The NaI(Tl) crystal has a limit on maximum allowable OH-impurity level before its properties are affected. However there are other 
materials allowing a large OH-content, which could be investigated. For example the Corning 7980 Fused silica has 800-1000~ppm of OH-content by weight.
One could perhaps consider enhancing the S/N ratio by implementing a notch filter to accept only wavelengths betwen 270~nm and 350~nm. 

\begin{figure}[tbp]
\includegraphics[width=0.5\textwidth]{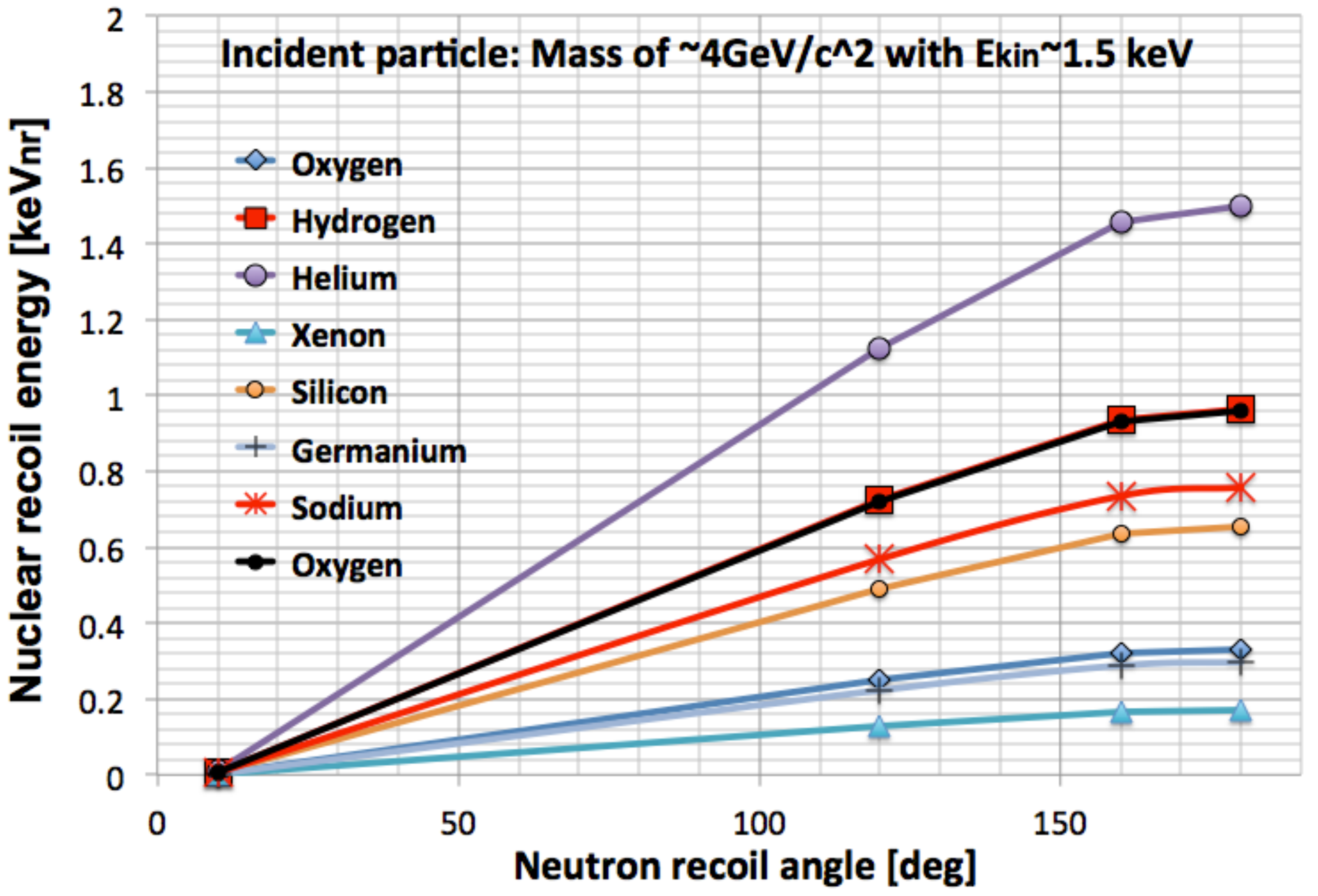}
\caption{Nuclear recoil energies for various nuclei assuming a $\sim$4~GeV/$c^2$ WIMP with kinetic energy of $\sim$1.5~keV.}
\label{fig:Nuclear_recoil_from_4GeV}
\end{figure}

Figure~\ref{fig:Nuclear_recoil_from_4GeV} shows nuclear recoil energies as a function of recoil angle for various nuclei, 
assuming the WIMP mass of 4~GeV/$c^2$ WIMP. One can see that the recoil energy for proton is about the same that from oxygen, i.e., the WIMP collision with either proton or oxygen 
can excite the OH-molecule. One should note that a recoil energy of sodium nucleus is below DAMA's threshold of $\sim$1.5keV.

\section{Suggested steps to calibration the DAMA crystals}

One has to show that the OH-molecule imbedded in NaI(Tl) crystals behave the same way as described in this paper.
To prove the OH-hypothesis one should expose DAMA crystals to a very low energy sub-keV neutron beam and check the crystal response. If this result is 
positive, one can parameterize a dependence on the OH-content in small specially prepared NaI(Tl) samples with different OH-content. In addition, one could also 
use the laser-induced fluorescence method to quantify the OH-content in some DAMA crystals. There are chemists specializing in this methodology.

We believe that these calibration steps are necessary to understand the NaI(Tl) crystal response to very low energy proton recoils. One should not use the energy 
dependence obtained form the Gamma source calibration, as it will likely produce incorrect conclusions of the NaI(Tl) crystal response from sub-keV deposits of low mass WIMPs.

\section{Conclusion}

This paper provides arguments why the OH-molecule may be a good way to detect very low mass WIMP. It provides a method to reach the lowest possible WIMP mass.
The paper suggests concrete steps the DAMA group could take to prove that the proposed idea is valid. 
The paper also suggests other ways to create a detector with large OH-content, which could be used for a detection of very low mass WIMPs.

\end{document}